\documentclass[aps,prb,twocolumn]{revtex4-1}
\usepackage{amssymb}
\usepackage{amsfonts}
\def\checkT{{\check{T^{\vphantom{'}}}} }

\newcommand{\be}{\begin{equation}}
\newcommand{\bd}{\begin{displaymath}}
\newcommand{\ee}{\end{equation}}
\newcommand{\ed}{\end{displaymath}}
\newcommand{\ba}{\begin{eqnarray}}
\newcommand{\ea}{\end{eqnarray}}

\newcommand{\bW}{{W}}

\newcommand{\half}{{\textstyle \frac{1}{2}}}
\newcommand{\halfs}{{\scriptstyle \frac{1}{2}}}
\newfont{\mycal}{eufb10 at 12pt}
\newfont{\myeu}{eufm10}
\newcommand{\e}{{\rm e}}

  \newcommand{\cac}{\mathfrak{c}}

\def\checkT{{\check{T^{\vphantom{'}}}} }

   \newcommand{\rd}{{\rm d}}

\begin{document}
\title[Layered Ising Model]%
{Criticality in Alternating Layered Ising Models: \\
II. Exact Scaling Theory}

\author{Helen Au-Yang}
\affiliation{Department of Physics, Oklahoma State University, %\\
145 Physical Sciences, Stillwater, OK 74078-3072, USA}
\email{helenperk@yahoo.com}

\begin{abstract}
Part I of this article studied the specific heats of planar
alternating layered Ising models with strips of strong coupling
$J_1$ sandwiched between strips of weak coupling $J_2$,
to illustrate qualitatively the effects of connectivity,
proximity, and enhancement in analogy to those seen in
extensive experiments on superfluid helium by Gasparini
and coworkers. It was demonstrated graphically that
finite-size scaling descriptions hold in a variety of
temperature regions including in the vicinity of the
two specific heat maxima. Here we provide exact theoretical
analyses and asymptotics of the specific heat that support
and confirm the graphical findings. Specifically, at the
overall or bulk critical point, the anticipated (and always
present) logarithmic singularity is shown to vanish exponentially
fast as the width of the stronger strips increases.
\end{abstract}
\maketitle

%%%%%%%%%%%%%%%%%%%%%%%%%%%%%%%%%%%%%%%%%%%%%%%%%%%%%%%%%%
The previous paper, Part I,\cite{HAYMEF} considered a range of
exactly soluble Alternating Layered Ising (ALI) models and presented
extensive plots of their specific heats. The primary motivation
(as explained in Part I) was to illustrate and study theoretically
the phase-transition phenomena of ``proximity,'' ``connectivity,''
and ``enhancement'' as highlighted experimentally by recent studies
of superfluid helium by Gasparini and
coworkers.\cite{GKMD,PKMG,PKMGn,MEF,PG,PKMGpr} However,
the ALI models have intrinsic interest as instructive examples
of the general two-dimensional layered Ising models. The exact
solubility of the general layered models was reported in
1969\cite{MEFJp} and noted and developed, independently, in the
context of randomly coupled systems by McCoy and Wu\cite{MWbk}
and further studied in Refs.~\onlinecite{HAYBM,Hamm}. 
 
Specifically, our work addresses ALI models in which, in a standard
infinite two-dimensional square lattice with Ising spins at each
site, $(i,j)$, infinite strips of width $m_1$ are coupled by nearest
neighbor (nn) energies of strength $J_1$ in alternation with infinite
strips of width $m_2=sm_1$ and coupling $J_2=rJ_1$. More explicitly,
the nn couplings between spins at sites $(i,j)$, $(i,j+1)$
and $(i+1,j)$ in the same strip are independent of $j$ but equal
to $J_1$ for $i=1,2,\cdots,m_1$ but to $J_2$ for
$i=m_1+1,m_1+2,\cdots,(m_1+m_2)$, and so on with overall period
$(m_1+m_2)$. The boundary spins separating layers are thus at
$i=1+n(m_1+m_2)$ and at $i=1+(n+1)m_1+nm_2$ for $n=0,\pm1,\pm2,\cdots$.

This paper then presents the details of the exact calculations on
which the plots (and discussion) of Part I was based. In Section I,
we present the specific integrals for the free energy without giving
detailed derivations, because the ALI models are special cases of
the general layered models\cite{HAYBM,Hamm} where the details can be
found. The explicit forms of the integrals were used to plot the
specific heat for the alternating layered systems shown in Figs.\ 2
and 3 of Part I. Around the overall or unique bulk critical point,
$T_c(r, s)$, the specific heat diverges logarithmically with an
amplitude, $A(r,s)$, that is shown in Section II to decrease
exponentially fast with increasing $m_1$. In Section III, we examine
the behavior of the free energy near $T_{1c}$ and $T_{2c}$, which are,
respectively, the bulk critical temperatures of uniform 2-D Ising
models with strong couplings, $J_1$, and weak couplings $J_2$, or,
otherwise, the limiting temperatures of the specific heat maxima
for infinitely wide layers. The conditions for the data collapse
shown in Figs.\ 4 and 5 of Part I are thus verified. In Section IV,
we show that finite-size scaling holds in both regimes, whenever
data collapse occurs. Finally, Section V studies the behaviors of
the enhancements in the two regimes, while the paper ends with a
short summary.

%%%%%%%%%%%%%%%%%%%%%%%%%%%%%%%%%%%%%%%%%%%%%%%%%%%%%%%%%%
\section {The Free Energy}
Layered Ising models were studied earlier in
Refs.~\onlinecite{MEFJp,MWbk,HAYBM,Hamm,KardarBerker}.
As the alternating layered model with cyclic boundary conditions
imposed in the infinite vertical direction, and free boundary
conditions in the horizontal layered direction are special cases,
the calculation for the free energy per site, ${f(J_1,J_2;m_1,m_2;T)}$,
is almost identical to that in Refs.~\onlinecite{HAYBM,Hamm}. Thus
for layers of thickness $m_1$ and couplings $J_1$ sandwiched between
layers of thickness $m_2$ and couplings $J_2$, one has
\ba 
&&-\frac {f(J_1,J_2;T)}{k_B T}
=\frac {\ln[(2S_1)^{m_1} (2S_2)^{m_2}]}{2(m_1+m_2)}\nonumber\\
&&\qquad+\frac 1{m_1+m_2}\int_0^{\halfs\pi}
\frac{\rd\theta}{\pi}\ln\half\Big[\bW+\sqrt{\bW^2-4}\Big], 
\label{dff}\ea
where, for $i=1,2$, we use here and below
\be
S_i=\sinh 2K_i,\quad C_i=\cosh 2K_i,
\ee
with
\be K_1=J_1/(k_BT),\quad K_2\equiv rK_1,
\ee
while the function $W(T;J_1,J_2;m_1,m_2;\theta)$ is given by
\ba
\bW=U^+_1U^+_2+U^-_1U^-_2+\half(C_1C_2-1)V_1V_2,
\label{dfW}\ea
in which for $i=1,2$,
\ba
U^{\pm}_i&=&U^{\pm}(t_i,m_i)\nonumber\\
&=&\half(\alpha_i^{m_i}+\alpha_i^{-m_i})
\pm\half(\alpha_i^{m_i}-\alpha_i^{-m_i})g_i,\nonumber\\
V_i&=&(\alpha_i^{m_i}-\alpha_i^{-m_i}){\bar g}_i,\hspace{1in}
\label{dfUV}\ea
where we have introduced the basic temperature variables, $t_i$, via
\ba 
t_i=\big(1-S_i\big)\Big/\sqrt{2S_i}\approx2K_{ic}-2K_i
\approx2K_{ic}(T/T_{ic}-1),\nonumber\\
2K_{ic}=\ln(\sqrt2+1),\hspace{1in}
\label{dft}\ea
which are identical to the variables used in
Refs.~\onlinecite{FerdFisher,HAYFisher}, but differ from symbols
$t_1$ and $t_2$, defined in (I.13) and (I.18) of Part I although
by only a constant factor when close to $T_{1c}$ and $T_{2c}$,
respectively. The amplitude functions in (\ref{dfUV}) are then
\ba
g_i&=&g_i(t_i;\omega)\nonumber\\
&=&\Big[t_i\sqrt{1+t_i^2}(1-\omega^2)
+\omega^2\sqrt{(1+t_i^2)(2+t_i^2)}\Big]\Big/Y_i,\nonumber\\
{\bar g}_i&=&{\bar g}_i(t_i;\omega)=\omega\sqrt{1-\omega^2}\Big/Y_i,
\nonumber\\
Y_i&=&Y(t_i;\omega)=\sqrt{(\omega^2+t_i^2)(1+\omega^2+t_i^2)},
\label{dfgi}\ea
while $\omega=\sin\theta$, and the layer spacings $m_1$ and $m_2$,
enter through
\ba 
\alpha_i&=&\alpha(t_i)={\cac}_i+\sqrt{{\cac}_i^2-1}, \nonumber\\
{\cac}_i&=&{\cac}(t_i)=2 t^2_i+2\omega^2+1.
\label{alpha}\ea
We remark that by comparing with equations (2.5) in
Ref.~\onlinecite{FerdFisher} one finds ${\cac}(t)=c_\ell$ with
$1-2\omega^2=\cos(\pi\ell/n)$. It is also easily seen that
$2Y_i=\sqrt{\cac_i^2-1}$. Conversely, we also have the relations
\ba 
C_i&=&\sqrt{1+t_i^2}\Big[\sqrt{2+t_i^2}-t_i\Big], \nonumber\\
S_i&=&1+ t^2_i-t_i\sqrt{2+t_i^2}.
\label{CS}\ea

The terms $U^{+}_i=U^{+}(t_i,m_i)$ in (\ref{dfW}) are related to
the free energy $f^{\infty}(m_i;J_i;T)$ of an infinite strip of
width $m_i$ with coupling energy $J_i$ which is\cite{HAYFisher}
\ba 
&&-\frac {f^{\infty}(m_i;J_i;T)}{k_B T}
=\frac { \ln(2S_i )}{2}\nonumber\\
&&\qquad+\frac 1{\pi m_i}\int_0^{\pi/2}\rd \theta \ln U^{+}(t_i,m_i).
\label{finites}\ea
The remaining terms in (\ref{dfW}) are related to the interaction
between the strips. If $J_2\to0$, so that the system becomes
uncoupled, the relations (\ref{dft}) yield
$t_2\to(2S_2)^{-1/2}\to\infty$, which is used in (\ref{alpha}) to
give $\alpha_2\to 4 t^2_i\to 2/S_2$. Consequently, from (\ref{dfUV})
we have $U^+_2=\halfs\alpha_2^{m_2}=2^{m_2-1}/(S_2)^{m_2}$, and
from (\ref{dfgi}) we find $g_2\to1$ and ${\bar g}_2\to0$. These
results establish $U^-_2=0$ and $V_2=0$. In this limit, the free
energy in (\ref{dff}) becomes
\ba 
&&-\frac {f(J_1,0;T)}{k_B T}
=\frac { m_1\ln(2S_1 )+2m_2\ln 2}{2(m_1+m_2)}\nonumber\\
&&\qquad+\frac 1{\pi(m_1+m_2)}
\int_0^{\pi/2}\rd\theta\ln U^{+}(t_1,m_1), 
\label{finite1}\ea
which is the free energy per site for infinite strips of width
$m_1$, coupling $J_1$, separated by empty infinite strips of
width $m_2$. This is identical to the result in (\ref{finites}),
except that the factor $1/m_1$ in (\ref{finites}) is replaced
by $1/(m_1+m_2)$, while the additional constant term,
$2m_2\ln2/(m_1+m_2)$, does not contribute to the specific heat.

For completeness, we also let $J_1=0$ to find
\ba 
&&-\frac {f(0,J_2;T)}{k_B T}
=\frac { m_2\ln(2S_2 )+2m_1\ln 2}{2(m_1+m_2)}\nonumber\\
&&\qquad+\frac 1{\pi(m_1+m_2)}
\int_0^{\pi/2}\rd\theta\ln U^+(t_2,m_2).
\label{finite2}\ea
For future purposes [entailed in establishing relations (I.19),
and (I.21)], we recall the modified temperature ${\checkT}(T)$,
introduced in (I.20), and then define 
\ba 
&&-\frac {f(0,J_2;{\checkT})}{k_B T}
=\frac { m_2\ln(2S_2 )+2m_1\ln 2}{2(m_1+m_2)}\nonumber\\
&&\qquad+\frac 1{\pi(m_1+m_2)}
\int_0^{\pi/2}\rd\theta\ln U^+(-t_2,m_2). 
\label{finite2c}\ea

Using (\ref{dfUV}), we may rewrite (\ref{dfW}) as
\ba
&&\bW=\half(\alpha_1^{m_1}+\alpha_1^{-m_1})(\alpha_2^{m_2}
+\alpha_2^{-m_2})\nonumber\\
&&\qquad+\half(\alpha_1^{m_1}-\alpha_1^{-m_1})(\alpha_2^{m_2}
-\alpha_2^{-m_2})G(t_1,t_2;\omega),
\label{dfWn}\ea
in which we have
\ba
(Y_1Y_2)G(t_1,t_2;\omega)=
(Y_1Y_2)[g_1g_2+(C_1C_2-1){\bar g}_1{\bar g}_2]\nonumber\\
=\bigg[t_1t_2\sqrt{(1+t_1^2)(1+t_2^2)}-
\omega^2\bigg](1-\omega^2)\qquad\nonumber\\
+\omega^2\sqrt{(1+t_1^2)(1+t_2^2)(2+t_1^2)(2+t_2^2)}.\qquad
\label{dfG}\ea

For the uniform Ising model the ratio $r=J_2/J_1$ is unity, so that
$t_1=t_2$ and $\alpha_1=\alpha_2$. We now use (\ref{dfgi}) and
(\ref{CS}) to show that $G$ in (\ref{dfG}) reduces to
$G=g_1^2+S_1^2{\bar g}_1^2=1$. As a result (\ref{dfWn}) simplifies to
\ba
\bW&=&\half(\alpha_1^{m_1}+\alpha_1^{-m_1})
(\alpha_1^{m_2}+\alpha_1^{-m_2})
\nonumber\\
&&+\half(\alpha_1^{m_1}-\alpha_1^{-m_1})
(\alpha_1^{m_2}-\alpha_1^{-m_2})\nonumber\\
&=&(\alpha_1^{m_1+m_2}+\alpha_1^{-m_1-m_2}).
\ea
Consequently, the free energy in (\ref{dff}) becomes 
\ba 
-\frac {f(J_1,J_1;T)}{k_B T}=\frac 1{2}\ln(2S_1) 
+\frac 1{\pi}\int_0^{\pi/2}\rd \theta\,\ln\alpha_1,
\label{dffu}\ea
which is the same as the free energy of the uniform Ising model.\cite{MWbk}

The specific heat of the alternating layered model, which is the
second derivative of the free energy in (\ref{dff}), is thus given by
\ba
&&\frac {C(J_1,J_2;m_1,m_2;T)}{k_B}=
K_1^2\frac{d^2}{dK_1^2}\bigg[-\frac{f(J_1,J_2;T)}{k_BT}\bigg]
\nonumber\\
&&\qquad=\frac{-2m_1 K_1^2}{(m_1+m_2)S^2_1}
-\frac{2m_2 (rK_1)^2}{(m_1+m_2)S^2_2}\nonumber\\
&&\quad+\frac{K_1^2}{\pi(m_1+m_2)}
\int_0^{\pi/2}\rd\theta\Bigg[\Bigg({\frac{d^2\bW}{dK_1^2}}\Bigg)
\bigg/\bigg(\bW^2-4\bigg)^{\frac 12}\nonumber\\
&&\qquad-\bW\Bigg(\frac {d\bW}{dK_1}\Bigg)^2\bigg/
\bigg(\bW^2-4\bigg)^{\frac 32}\Bigg].\,\,
\label{spheat}\ea
These are the formulae used to plot the specific heats in the
figures in the previous paper, Part I.\cite{HAYMEF}

In considering the expression (I.1) for the specific heat near the
bulk critical point $T_c$, it is natural, having dealt with the
amplitude, $A(r, s)$, of the logarithmic singularity, to inquire as
to the leading background term, $B(r, s)$. As our notation suggests,
this is expected, on the grounds of duality,\cite{MKauf} to be
continuous through $T_c$ so that there is no discontinuity associated
with bulk criticality. However, the calculation of the dependence of
$B(r, s)$ on $m_1$ proves not straightforward and has not been
attempted (although the continuity is surely supported by the
numerics reported in Part I).

%%%%%%%%%%%%%%%%%%%%%%%%%%%%%%%%%%%%%%%%%%%%%%%%%%%%%%%%%%
\section {Amplitude for the Logarithmic Divergence}
The amplitude of the logarithmic divergence in (I.1) is
obtained\cite{HAYBM,Hamm} by expanding the term inside the
square root in (\ref{dff}) as
\ba
&&1-4/W^2=\nonumber\\ 
&&\quad A_1^2(J_1/k_B)^2[(1/T)-(1/T_c)]^2+A^2_2\theta^2+\ldots,
\ea
where the coefficient $A_1$ is given by Hamm in (1.8) of
Ref.~\onlinecite{Hamm} as
\ba
A_1&=&2m_1(1+S^{-1}_{1c})+2m_2r(1+S^{-1}_{2c}),\nonumber\\
S_{1c}&=&\sinh 2K_c,\quad S_{2c}=\sinh (2rK_c).
\label{A1}\ea
The integration over $\theta$ around the origin yields
\ba C(T)/k_B&=&-A(r,s)\ln|1-(T/T_c)|+\mathrm{O}(1),\nonumber\\
&&\quad r=J_2/J_1,\quad s=m_2/m_1,\nonumber\\
A(r,s)&=&\frac{A_1^2K_c^2}{2\pi A_2(m_1+m_2)},\quad
K_c=\frac{J_1}{k_BT_c}.
\label{log}\ea

Since we only have two kinds of bonds, the sum in (1.9) of
Ref.~\onlinecite{Hamm} can be evaluated to obtain
\ba
A_2^2&=&(\epsilon_{1c}^{m_1}-\epsilon_{1c}^{-m_1})^2\Bigg[
\frac{S_{1c}^2}{(\epsilon_{1c}-\epsilon^{-1}_{1c})^2}+
\frac{S_{2c}^2}{(\epsilon_{2c}-\epsilon^{-1}_{2c})^2}
\nonumber\\
&&\qquad-\frac{S_{1c}S_{2c}(z_{1c}z^{-1}_{2c}+z_{2c}z^{-1}_{1c})}
{(\epsilon_{1c}-\epsilon_{1c}^{-1})
(\epsilon_{2c}-\epsilon_{2c}^{-1})}\Bigg],
\label{A2}\ea
where the temperature dependent parameters are
\ba
z_{i}=z_{i}(T)&=&\tanh (J_i/k_BT), \nonumber\\ 
\epsilon_{i}(T)&=&z_{i}\e^{2(J_i/k_BT)},
\label{epsilon}\ea
while $z_{ic}=z_{i}(T_c)$ and $\epsilon_{ic}=\epsilon_{i}(T_c)$.
Notice that $\epsilon_{i}(T)$ depends only on $J_i$, and at the
critical temperature $T_{ic}$ of a uniform planar Ising model whose
coupling energy is $J_i$, we have
\be
\epsilon_{i}(T_{ic})=1 \Rightarrow K_{ic}=J_i/(k_BT_{ic})=
\halfs\ln(\sqrt{2}+1),
\label{tc12}\ee 
which is equivalent to (I.5).\cite{HAYMEF} The general critical
temperature expression (I.4) is equivalent to
\be
\epsilon^{m_1}_{1}(T_c)\epsilon^{m_2}_{2}(T_c)=1,\quad \hbox{or}\quad
\epsilon_{1c}\epsilon^s_{2c}=1.
\label{tc}\ee
From (\ref{tc12}) and (\ref{tc}), we find $T_c<T_{1c}$, and
$T_c\to T_{1c}$ either as $s\to 0$ or as $r\to 1$. For $r\ne1$
and $s\ne0$, we find $\epsilon_{1c}>1$ and $\epsilon_{2c}<1$, so
that $\epsilon_{1c}^{m_1}=\epsilon_{2c}^{-m_2}\to \infty$ in the
limit $m_1\to\infty$. We shall consider the scaling behavior for
the two cases separately.

\begin{itemize}
\item{}Now consider the scaling limit for $r\ne1$, $m_2=s m_1$ fixed,
and $m_1\to \infty$ such that $s\to0$. We find from (\ref{tc12})
and (\ref{tc})
\ba
\ln\epsilon_1(T_c)-\ln\epsilon_1(T_{1c})&=&-s\ln\epsilon_2(T_c)
\nonumber\\ 
&\approx& -s\ln\epsilon_2(T_{1c}).
\ea
Now we substitute (\ref{epsilon}) into this relation to find
\ba
\frac{2J_1}{k_B}\Bigg[\frac1{T_c}-\frac1{T_{1c}}\Bigg]+
\ln\Bigg[\frac{\tanh(J_1/k_BT_{c})}{ \tanh(J_1/k_BT_{1c})}\Bigg]
\nonumber\\
\approx
s\ln\Bigg[\frac{1+\e^{-2rJ_1/k_BT_{1c}}}{\e^{2rJ_1/k_BT_{1c}}-1}\Bigg].
\label{tcs0}\ea
After expanding the left hand-side around $T_{1c}$ and using
(\ref{tc12}) on the right we find
\ba
&&4K_{1c}[(T_{1c}/T_c)-1]\approx s\cdot p_r,\nonumber\\
&&p_r=\ln[(\sqrt2-1)^r+1]-\ln[(\sqrt2+1)^r-1].
\ea

Consequently for $s\to 0$, we have
\ba
A_1&=&4m_1[1+\mathrm{O}(s)],\qquad\nonumber\\
A_2&=&\frac{\epsilon^{m_1}_{1c}-\epsilon^{-m_1}_{1c}}
{\epsilon_{1c}-\epsilon^{-1}_{1c}}+\mathrm{O}(1)\approx
\frac{\sinh m_2 p_r}{s p_r}.
\ea
The amplitude of the logarithmic divergence scales as
\be A(r,s)\approx\frac{8K_{1c}^2 p_r m_2}{\pi\sinh (p_r m_2)}
+\mathrm{O}(s).
\label{scaling}\ee
For $r=1$, we have $p_r=0$, which reproduces the original Onsager
result.\cite{MWbk, FerdFisher, HAYFisher} For $m_2\to0$, we find
$\sinh(p_r m_2)\to p_r m_2$, so that (\ref{scaling}) again reproduces
the Onsager result. In the opposite limit $m_2\to\infty$, we find the
amplitude decays exponentially fast as
\be
A(r,s)=({8K_{1c}^2 }/{\pi} )\cdot p_r m_2\,\e ^{-p_r m_2}.
\label{amp1}\ee

\item{}In order to have a non-vanishing logarithmic amplitude
for fixed $s=m_2/m_1$, with $m_1\to\infty$, one must let $r\to 1$.
Accordingly, we study the amplitude of the logarithmic
singularity in the scaling limit that $(1-r)m_1$ is fixed.
In similar fashion to our derivation of (\ref{tcs0}), we use
(\ref{tc12}) and (\ref{tc}) to find
\ba
\frac 1{T_c}&\approx&\frac 1{T_{1c}}\Bigg[1+\frac{s(1-r)}{1+s}\Bigg],\nonumber\\ 
\frac r{T_c}&\approx&\frac 1{T_{1c}}\Bigg[1-\frac{1-r}{1+s}\Bigg].
\ea
Expanding terms in (\ref{A1}) and (\ref{A2}) as a series in $1-r$,
and keeping only the leading two terms, we obtain
\ba
S_{1c}&\approx&1+\frac{2\sqrt{2}K_{1c}s(1-r)}{1+s},\nonumber\\
S_{2c}&\approx&1-\frac{2\sqrt{2}K_{1c}(1-r)}{1+s},\nonumber\\
z_{1c}/ z_{2c}&\approx&1+2K_{1c}(1-r),\nonumber\\
z_{2c}/ z_{1c}&\approx&1-2K_{1c}(1-r),\nonumber\\
\epsilon_{1c}&\approx&1+{4K_{1c}s(1-r)}/(1+s)\nonumber\\
&\approx&\e^{{4K_{1c}s(1-r)}/(1+s)},\nonumber\\
\epsilon_{2c}&\approx&1-{4K_{1c}(1-r)}/(1+s)\nonumber\\
&\approx&\e^{-{4K_{1c}(1-r)}/(1+s)}.
\label{exdelta}\ea
Substituting these asymptotic relations into (\ref{A1})
and (\ref{A2}), we find
\ba
A_1&\approx&4m_1(1+s),\qquad \nonumber\\
A_2&\approx&\frac{(1+s)^2}{4K_{1c}s(1-r)}
\sinh\left[\frac{4K_{1c}sm_1 (1-r)}{(1+s)}\right].
\ea
Consequently, the scaling form of the amplitude of the logarithmic
divergence of the specific heat is
\ba
A(r,s)&\approx&\frac{16K_{1c}^2sq}{\pi(s+1)\sinh[2sq/(1+s)]},
\nonumber\\
&&\qquad q=2K_{1c}(1-r)m_1.
\label{ascaling}\ea
For $r=1$, one has $q=0$, and this expression again reproduces the
original Onsager result. For $r<1$, with $m_1\to \infty$, we have
$q\gg1$ so the denominator is exponentially large, which means the
amplitude is exponentially small. This and (\ref{amp1}) are central
results that explain why the logarithmic singularity at $T_c$
becomes essentially unobservable in Figs.\ 2-4 of Part I.
\end{itemize}

%%%%%%%%%%%%%%%%%%%%%%%%%%%%%%%%%%%%%%%%%%%%%%%%%%%%%%%%%%
\section {Behavior near $T_{1c}$ and $T_{2c}$}
Near $T_{1c}$, the specific heat of the weaker strip is small,
and so in paper I we introduced a net contribution
\be
{C_1}(J_1,J_2;T)=(1+s)[C(J_1,J_2;T)-C(0,J_2;T)].
\label{C1}\ee
In Fig.\ 5 of Part I, \cite{HAYMEF} the scaling plots of $C_1$
reveal that the behavior becomes independent of $m_2$ for $T$
near $T_{1c}$. In this section, we examine the condition for such
behavior to hold. We define the free energy corresponding to
$C_1(J_1,J_2;T)$ as
\ba
{f_1}(J_1,J_2;T)=(1+s)[f(J_1,J_2;T)-f(0,J_2;T)].\cr
%{\check f}_1(J_1,J_2;T)=(1+s)[f(J_1,J_2;T)-f(0,J_2;{\check T})],
\label{f1}\ea
For fixed weakness ratio $r=J_2/J_1$, we can see that from
(\ref{alpha}) and (\ref{dft}) that $\alpha_1\simeq1$ for
$t_1,\omega\simeq 0$. Because $r\ne1$, for $T\simeq T_{1c}$,
($t_1\simeq 0$), we have $t_2\ne 0$ and $\alpha_2>1$. Thus for
$m_2$ sufficiently large, $\alpha_2^{m_2}\gg\alpha_2^{-m_2}$, and
we may drop terms involving $\alpha_2^{-m_2}$ in (\ref{dfWn}) to
arrive at the form
\ba
W&\approx&\alpha_2^{m_2}W_1(t_1,t_2;m_1)=
\alpha_2^{m_2}\Big[\half(\alpha_1^{m_1}+\alpha_1^{-m_1})\nonumber\\
&&\qquad+\half(\alpha_i^{m_1}-\alpha_1^{-m_1})G(t_1,t_2;\omega)\Big],
\label{dfW1}\ea
in which $G(t_1,t_2;\omega)$ was defined in (\ref{dfG}).
Similarly we find
\be
U^+(\pm t_2,m_2)=\alpha_2^{m_2}\Big[1+g_2(\pm t_2)\Big]+
\mathrm{O}(\alpha_2^{-m_2}).
\ee
Consequently the free energy introduced in (\ref{f1}) becomes
\ba
&&-{ f_1(J_1,J_2;T)}/{(k_BT)}
=\half \ln({S_1}/2)\hspace{1in}\nonumber\\
&&\qquad+(\pi m_1)^{-1}\int_0^{\halfs\pi}{\rd\theta}\,
[{\cal I}_1(t_1,t_2;m_1)+\mathrm{O}(\alpha_2^{-m_2})],\nonumber\\
&&{\cal I}_1(t_1,t_2;m_1)=\ln W_1(t_1,t_2;m_1)-
\ln\half[1+g_2(t_2)].\nonumber\\
\label{ftc1}\ea
It is easy to see from this that $ f_1$ is essentially independent
of $m_2$. Therefore, in the temperature range where
$\alpha_2^{m_2}\gg\alpha_2^{-m_2}$, which is the case for those
plots in Fig.~5 of Part I, the specific heat difference $C_1(T)$
defined in (\ref{C1}) becomes, indeed, independent of $m_2$. 

To understand the behavior of the integral in (\ref{ftc1}), we
compare it with the free energy of uncoupled strips when the
horizontal coupling, say ${\bar J}_2$, is set to be identically
zero in the weaker strips while the vertical couplings $J_2$
remain nonzero. Since the original papers\cite{HAYBM,Hamm} on
layered Ising systems address the more general case with vertical
couplings unequal to the horizontal couplings, we may easily obtain
the free energy for such a case as
\ba
&&-{ f^{(u)}(J_1;J_2;T)}/({k_BT})=
[{2(m_1+m_2)}]^{-1} \Bigg[m_1\ln(2S_1)\nonumber\\
&&\qquad+m_2[\ln2+\ln(1+C_2)]
+\frac 2{\pi }\int_0^{\halfs\pi}{\rd\theta}\ln W^{(u)}\Bigg],
\label{fu}\ea
where, in the integrand, we now have
\ba
&&W^{(u)}=\half(\alpha_1^{m_1}+\alpha_1^{-m_1})+
\half(\alpha_1^{m_1}-\alpha_1^{-m_1})g^{(u)}(t_1;J_2),\nonumber\\
&&g^{(u)}(t_1;J_2;\omega)=g_1(t_1)+\nonumber\\
&&\frac{\omega^2(1-\omega^2)[-t_1\sqrt{1+t_1^2}+\sqrt{(1+t_1^2)(2+t_1^2)}-1]}{(\omega^2-\half-\half C_2/S_2)Y_1},
\label{gu}\ea
where we recall that $Y_1=Y(t_1;\omega)$ is defined in (\ref{dfgi}).
In ${\cal I}_1(t_1,t_2;m_1)$ of (\ref{ftc1}), the second term is
related to the surface free energy, while the first term is very
similar to the free energy of uncoupled infinite strips of width
$m_1$, each with one of its boundary columns having vertical
couplings $J_2$. This can be seen for $t_2>0$ by rewriting the
function $G$ given by (\ref{dfG}) as
\ba
&&G(t_1,t_2;\omega)=g_1(t_1)\nonumber\\
&&\quad+{\omega^2(1-\omega^2)}{Y_1}^{-1}
\Big[-{t_1\sqrt{(1+t_1^2)}}R_1(t_2;\omega)\nonumber\\
&&\qquad+{\sqrt{(1+t_1^2)(2+t_1^2)}}R_2(t_2;\omega)-{Y_2}^{-1}\Big],
\label{Gt1}\ea
in which we have
\ba
&&R_1(t_2;\omega)=
\frac{1+2t_2^2+\omega^2}{Y_2(Y_2+t_2\sqrt{1+t_2^2})},\nonumber\\
&&R_2(t_2;\omega)=
\frac{2+2t_2^2+\omega^2}{Y_2[Y_2+\sqrt{(1+t_2^2)(2+t_2^2)}]}.
\label{R12}\ea
When $r\ne1$, for $t_1\simeq0$ we find that $t_2$ is large and
positive, so that $R_1(t_2;\omega)$ and $R_2(t_2;\omega)$ are not
singular. It is easily seen from (\ref{Gt1}) and (\ref{gu}) that
though the functions $G$ and $g^{(u)}$ are different, yet both
differ from $g_1(t)$ by factors which are of the order
$\omega^2/Y_1$, which do not contribute to the scaling function
as shall be shown later in Sect.~IV.

Similarly, for $T\simeq T_{2c}$, so that $\alpha_2\simeq 1$,
but $\alpha_1>1$, we see that whenever
$\alpha_1^{m_1}\gg\alpha_1^{-m_1}$, we may drop the terms
$\alpha_1^{-m_1}$ entering $W$ in (\ref{dfWn}), to find
\ba
&&W=\alpha_1^{m_1}W_2(t_1,t_2;m_2)
+\mathrm{O}(\alpha_1^{-m_1}),\nonumber\\ 
&&W_2(t_1,t_2;m_2)=\half(\alpha_2^{m_2}+\alpha_2^{-m_2})
\nonumber\\
&&\qquad+\half(\alpha_2^{m_2}-\alpha_2^{-m_2})G(t_1,t_2).\nonumber\\
&&U^+(t_1,m_1)=\alpha_1^{m_1}\Big[1+g_1(t_1)\Big]
+\mathrm{O}(\alpha_1^{-m_1}).
\label{dfW2}\ea
As a consequence we find from (\ref{dff}) and (\ref{finite2}) that
\ba
-&&\frac{ f_2(J_1,J_2;T)}{k_BT}=\hspace{1.5in}\nonumber\\
&&\qquad\frac {m_1+m_2}{ m_2 k_B T}[-f(J_1,J_2;T)+f(J_1,0;T)]
\label{f2}\\
&&\quad=\half\ln(S_2/2)+
\int_0^{\halfs\pi}\frac{\rd\theta}{\pi m_2}\Big [\ln W_2(t_1,t_2;m_2)
\nonumber\\
&&\qquad-\ln\half(1+g_1)+\mathrm{O}(\alpha_1^{-m_1})\Big].
\label{ftc2}\ea
Since $ f_2$ is independent of $m_1$, the plots of $C_2(T)$ in
Fig.~6 of Part I\cite{HAYMEF} for different $m_1$ lie on the same
curve demonstrating the data collapse. 

From (\ref{alpha}), we find for $\omega\sim0$ the results
\ba
\alpha_1^{-m_1}&\approx& \e^{-2|t_1|m_1}\propto \e^{-2m_1/\xi_1(T)},
\nonumber\\
\alpha_2^{-m_2}&\approx& \e^{-2|t_2|m_2}\propto \e^{-2m_2/\xi_2(T)},
\ea
where $\xi_i(T)$ is the bulk correlation length of the uniform Ising
model with couplings $J_i$. This means that if $r$ increases, so that
$t_2$ becomes closer to $ t_1$, then for (\ref{dfW1}) to hold, so
that data collapse occurs as shown in Fig.\ 5 of Part I,\cite{HAYMEF}
we must have $m_2$ large. Likewise as $r$ increases, one sees that
relations (\ref{dfW2}) still are valid provided $m_1$ is large with
the consequence that data collapse still occurs near $T_{2c}$.

Even though (\ref{dfW2}) in (\ref{ftc2}) looks similar to
(\ref{dfW1}) in (\ref{ftc1}), there are significant differences.
In the regime, $T\simeq T_{1c}$, the deviation $t_2$ is large and
positive, while for $T\simeq T_{2c}$, one finds that $t_1$ is a
large negative number, so that instead of (\ref{Gt1}), $G$ in
(\ref{dfG}) for $T\simeq T_{2c}$ behaves as
\ba
&&G(t_1,t_2;\omega)=g_2(-t_2)\hspace{1.2in}\nonumber\\
&&\qquad+\omega^2(1-\omega^2){Y_2}^{-1}
\Big[{t_2\sqrt{(1+t_2^2)}}R_1(|t_1|;\omega)\nonumber\\
&&\qquad\quad+{\sqrt{(1+t_2^2)(2+t_2^2)}}
R_2(t_1;\omega)-{Y_1}^{-1}\Big],
\label{Gt2}\ea
where $R_1(|t_1|;\omega)$ and $R_2(t_1;\omega)$ are seen from
(\ref{R12}) to be nonsingular for $t_1$ large.
Comparing this relation with (\ref{Gt1}), the flipping of the
sign of $t_2$ in $g_2$ is the reason that the rounded peak at
$T_{1max}$ is {\it below} $T_{1c}$, while $T_{2max}$ is {\it above}
$T_{2c}$. This then sets the stage for what otherwise might be
regarded as a purely phenomenological introduction of the modified
temperature variable ${\checkT}(T)=T_{2c}-(T-T_{2c})$ in (I.20).
 
%%%%%%%%%%%%%%%%%%%%%%%%%%%%%%%%%%%%%%%%%%%%%%%%%%%%%%%%%%
\section {Scaling functions}
We now consider $f_1$ in (\ref{ftc1}) in the scaling limit
$m_1\to\infty$ and $T\to T_{1c}$, and show that its scaling function
is identical to that in (\ref{finites}) for a infinite strip of
width $m_1$ and couplings $J_1$.\cite{HAYFisher} In fact we shall
show that when the differences in the integrands are of the order
$\omega^2/Y_1$, as in (\ref{Gt1}) or in (\ref{gu}), the scaling
functions remains unchanged. We shall outline now the steps used to
obtain the scaling function. 
\begin{itemize}
\item{Step 1:} We first change the integration variable in
(\ref{ftc1}) to $\omega=\sin\theta$, and then split the interval
of integration over $\omega$ into two parts, namely
$[0,1]\to[0,c/m_1]+[c/m_1,1]$, where here and below we take
$c=\ln m_1$. Then we will approximate the integrand differently
in the two distinct intervals.

\item{Step 2:} In the interval $[0,c/m_1]$, $\omega$ and $t_1$
are small, so we make the approximation
\ba
&&g_2\approx1,\quad\alpha_1^{m_1}=
\e^{2m_1\arcsin\sqrt{t_1^2+\omega^2}}\approx\e^{2X_1},\nonumber\\
&& X_1=\sqrt{\tau_1^2+\phi^2},\quad \tau_1=m_1t_1,\quad
\phi=m_1\omega,\nonumber\\
&&G(t_1,t_2)=g_1+\mathrm{O}(\omega^2)\nonumber\\
&&\qquad\qquad\approx { t_1}\Big/{ \sqrt{t_1^2+\omega^2}}
={\tau_1}/{X_1},
\label{appro}\ea
Note especially the introduction of the scaling variable $\tau_1$;
this is used in order to conform to the convention of the previous
papers \cite{FerdFisher, HAYFisher} in place of the scaling variable
$x_1$ used in Part I.\cite{HAYMEF} But, as seen from (\ref{dft}),
the two variables are related simply by a constant, i.e.,
$\tau_1=2K_{1c}x_1$ with $2K_{1c}=\ln(\sqrt 2+1)$.
Using (\ref{appro}), the integrand in (\ref{ftc1}) can now be
written as
\ba
{\cal I}_1&\approx& \ln W_1\approx{\cal H}(\tau_1,\phi)\nonumber\\
&=&\ln[\cosh2X_1+\sinh2X_1({\tau_1}/{X_1})].
\label{integrand1}\ea
After changing the variable of integration from
$\omega\to\phi=m_1\omega$, we split the interval of integration
of $\phi$ to : $[0,c]=[0,1]+[1,c]$.
\item{Step 3:} In the interval $\omega\in[c/m_1,1]$, we find, in
(\ref{dfW1}), $\alpha_1^{m_1}\gg\alpha_1^{-m_1}$, so that
$\alpha_1^{-m_1}$ can be dropped in $W_1$, and the integrand in
(\ref{ftc1}) becomes
\ba 
{\cal I}_1(\omega)&\approx&
\ln[\alpha_1^{m_1}\half(1+G(t_1,t_2;\omega))]\nonumber\\
&&\qquad-\ln\half[1+g_2(t_2;\omega)].
\label{integrand2}\ea
\item{Step 4:} The integration over $\omega$ in the interval for
the integrand in (\ref{integrand2}) is then split into two parts
$[c/m_1,1]=[1/m_1,1]-[1/m_1,c/m_1]$. We denote the integrals over
$[1/m_1,1]$ by
\ba
{\Sigma}_1&=&\frac 1{\pi}
\int_{1/m_1}^1\frac{\rd\omega}{\sqrt{1-\omega^2}}\ln\alpha_1,
\nonumber\\
{\Omega}_1&=&\frac 1{\pi m_1}
\int_{1/m_1}^1\frac{\rd\omega}{\sqrt{1-\omega^2}}[\ln(1+G)
\nonumber\\
&&\qquad-\ln(1+g_2)].
\label{Omega} \ea 
In the interval $\omega\in[1/m_1,c/m_1]$, we use (\ref{appro})
for the integrand in (\ref{integrand2}), so that
\be 
{\cal I}_1\approx {\cal H'}(\tau_1,\phi) =
\ln[\e^{2X_1}\half(1+{\tau_1}/X_1)].
\ee
Changing the variable of integration $\omega\to\phi=\omega m_1$,
and combining it with the integral over the interval $[1,c]$ of the
integrand in (\ref{integrand1}) in step 2, we obtain 
\ba
\delta{\cal H}(\tau_1,\phi)&=&{\cal H}(\tau_1,\phi)-{\cal H'}(\tau_1,\phi)\quad\nonumber\\
&=&\ln[1+\e^{-2X_1}(X_1-{\tau_1})/({X_1+{\tau_1}})].
\label{dH}\ea
For $\phi\ge c$, we find
\be
\delta{\cal H}(\tau_1,\phi)\approx
\e^{-2X_1}(X_1-{\tau_1})/({X_1+{\tau_1}})\ll1.
\ee
Thus, the interval of integration $[1,c]$ can be extended to
$[1,\infty]$ with negligible error.
\item{Step 5:} Combining all the steps, we find
\be 
f_1(J_1,J_2;T)\approx{\cal F}_1(\tau_1)+{\Sigma}_1+{\Omega}_1,
\label{fso1}
\ee
where with ${\cal H}(\tau_1,\phi)$ defined in (\ref{integrand1})
and $\delta{\cal H}(\tau_1,\phi)$ defined in (\ref{dH}), we have
\ba 
&&{\cal F}_1(\tau_1)=\nonumber\\
&&\quad\frac 1{m_1^2\pi}\left[ \int_0^1{\rd \phi}\,
{\cal H}(\tau_1,\phi)
+\int_1^\infty{\rd \phi}\, \delta{\cal H}(\tau_1,\phi)\right].
\ea 
From (\ref{Gt1}), we find that for $\omega\in[0,c/m_1]$ or
$[1/m_1,c/m_1]$, the terms of the order of $\omega^2/Y_1$ may
be dropped; thence the scaling function for an infinite strip of
finite width in (\ref{finites}) and the scaling function for
(\ref{ftc1}) can differ only through the term in ${\Omega}_1$
introduced in (\ref{Omega}). Since $T\simeq T_{1c}$, we find that
$g_2(\pm t_2)$ is not singular, while its contribution is of
order $1/m_1$; hence it does not contribute to the scaling function.
\item{Step 6:} The integrals for the derivatives of $\Sigma_1$ in
(\ref{Omega}) can be calculated explicitly. After keeping only the
scaling terms we find
\ba
&&K_1^2\frac {d^2{\Sigma}_1}{dK_1^2}=\nonumber\\
&&\frac{8K_1^2}{\pi}
\int_{\frac1{m_1}}^1\frac{\rd\omega}{\sqrt{1-\omega^2}} 
\frac{\omega^2(1+\omega^2+t_1^2)-t_1^2(t^2+\omega^2)}
{[(1+\omega^2+t_1^2)(t^2+\omega^2)]^{3/2}}\nonumber\\
&&=({8K_{1c}^2}/{\pi})\Bigg[\ln m_1+{\textstyle\frac32}\ln2
-\ln\Big(1+\sqrt{\tau_1^2+1}\Big)
\nonumber\\
&&\hspace{1in} -1+\frac1{\sqrt{\tau_1^2+1}}\Bigg]+
\mathrm{O}\Big(\frac{\ln m_1}{m_1}\Big).
\label{sigma1}\ea
The explicit calculation of the second derivatives of $\Omega_1$
is very messy. However, it is easy to see that only the lower
limit of the integration at $1/m_1$ can contribute to the scaling
function. For $\omega\sim1/m_1$, the integrand can be expanded as
a series in terms of $t_1$ and $\omega$ with the results, on keeping
only the leading terms,
\ba
&&K_1^2\frac {d^2{\Omega}_1}{dK_1^2}\approx \frac{- K_1^2}{\pi m_1}
\int_{\frac 1{m_1}}^1 {\rd \omega}\Bigg[
\frac{4t_1}{(t_1^2+\omega^2)^{3/2}}\nonumber\\
&&\hspace{0.8in}+\frac{4}{t_1^2+\omega^2}-
\frac{8t_1^2}{(t^2+\omega^2)^2}\Bigg]\nonumber\\
&&\quad\approx- \frac{4K_{1c}^2}{\pi}\left[\frac1{\tau_1}
\Bigg(1-\frac 1{\sqrt{\tau_1^2+1}}\Bigg)+\frac1{1+\tau_1^2}\right].
\label{omega1}\ea
As a cross-check, we have also verified that this agrees with
the tedious explicit calculations. As the difference between $G$
and $g_1$ are of the order of $\omega^2/Y_1$, we find that by
replacing $G$ by $g_1$ in $\Omega_1$ does not change the scaling
function. This means that near $T_{1c}$, the net specific heat
$C_1(J_1,J_2; T)$ defined in (I.11) has the
{\it same scaling behavior} as an infinite strip of width $m_1$
and couplings $J_1$. Specifically, we find
\ba
&&C_1(J_1,J_2;T)\approx A_0\ln m_1+Q(\tau_1)\approx C^\infty(J_1;T), \nonumber\\
&&A_0={8K_{1c}^2}/{\pi}=2[\ln(\sqrt2+1)]^2/\pi, 
\label{uppersc}\ea
where
\ba
Q(\tau_1)&=&\half A_0\Bigg[ \int_0^1{\rd \phi}\,
\frac{d^2}{d\tau_1^2} {\cal H}(\tau_1,\phi)\nonumber\\
&&+\int_1^\infty{\rd\phi}\,\frac {d^2}{d\tau_1^2}
\delta{\cal H}(\tau_1,\phi)\nonumber\\
&&+3\ln2-2\ln\Big(1+\sqrt{\tau_1^2+1}\Big)\nonumber\\
&&-\Big(2+\frac 1{\tau_1}\Big)
\Big(1-\frac1{\sqrt{\tau_1^2+1}}\Big)-\frac1{1+\tau_1^2}\Bigg].
\ea
Letting $\sigma=0$ in (2.62) of Ref..~\onlinecite{HAYFisher} we
find that the scaling function given there is almost identical
to this result; however, the difference term, $-A_0\pi/4$ in
(2.62) turns out to be a slip.\cite{note1} More recently the
finite-size scaling functions for the Ising model have been
shown\cite{WuHuIzmailian} to be of universal character.
\end{itemize}

Now to study the specific heat near the lower special region
$T\simeq T_{2c}$, we may use the same steps to analyze the integral
in (\ref{ftc2}). Because of (\ref{Gt2}),
for $\phi=m_2\omega\in[0,1]$, we find that (\ref{dfW2}) becomes
\be
W_2\approx{\cal H}(-\tau_2,\phi)=
\ln[\cosh2X_2-\sinh2X_2({\tau_2}/{X_2})],
\ee
where
\be
\tau_2=m_2 t_2,\quad X_2=\sqrt{\tau_2^2+\phi^2};
\ee
for $\phi\in[1,\infty]$, the integrand is approximated by
\be
\delta{\cal H}(-\tau_2,\phi)=
\ln[1+\e^{-2X_2}(X_2+{\tau_2})/({X_2-{\tau_2}})].
\ee
Consequently, the integral in (\ref{ftc2}) becomes
\be
f_2(J_1,J_2;T)\approx{\cal F}_2(-\tau_2)+{\Sigma}_2+{\Omega}_2,
\ee
where
\ba
&&{\cal F}_2(-\tau_2)=
\frac 1{m_2^2\pi}\Bigg[\int_0^1{\rd\phi}\,{\cal H}(-\tau_2,\phi)
\nonumber\\
&&\hspace{1.0in}+\int_1^\infty{\rd \phi}\,
\delta{\cal H}(-\tau_2,\phi)\Bigg],\\
&&{\Sigma}_2=\frac 1{\pi}
\int_{1/m_2}^1\frac{\rd\omega}{\sqrt{1-\omega^2}}\ln\alpha_2,
\hspace{0.5in}\\
&&{\Omega}_2=\frac 1{\pi m_2}\int_{\frac 1{m_2}}^1
\frac{\rd\omega}{\sqrt{1-\omega^2}}[\ln(1+G)-\ln(1+g_1)].\qquad
\label{fso2}\ea 
Again the derivatives of ${\Sigma}_2$ and ${\Omega}_2$ can be
evaluated, with results which can be obtained from (\ref{sigma1})
and (\ref{omega1}) by replacing $\tau_1$ by $-\tau_2$, and $m_1$
by $m_2$. The second derivative of ${\cal F}_2(-\tau_2)$ can also
be evaluated to find\cite{note2} for $T\sim T_{2c}$
\be
C_2(J_1,J_2;T)\approx A_0\ln m_2+Q(-\tau_2), \quad \tau_2=t_2m_2.
\ee
Finally, comparing the free energy in (\ref{finite2c}) with $U^+$
given by (\ref{dfUV}) with (\ref{f2}) with $W_2$ given in
(\ref{dfW2}), and then using (\ref{Gt2}), we find that for
$T\sim T_{2c}$
\ba
(1+s^{-1})C(0,J_2;{\checkT})&&=C_2(J_1,J_2;T)+
\mathrm{O}\Big({\ln m_2}/{m_2}\Big)\nonumber\\
&&\approx A_0\ln m_2+Q(-\tau_2),
\label{lowersc}\ea
where ${\checkT}$ is defined in (I.19) relating to $t_2\to-t_2$.

%%%%%%%%%%%%%%%%%%%%%%%%%%%%%%%%%%%%%%%%%%%%%%%%%%%%%%%%%%
\section{Enhancement}
Since the lower maxima of the specific heats $C_2(J_1,J_2;T)$ of
the coupled system are above $T_{2c}$, while the maxima of the
specific heats $C(0,J_2;T)$ of the uncoupled system are
below $T_{2c}$, we have introduced in the specific heats
$C(0,J_2;{\checkT})$ whose free energy is defined in
(\ref{finite2c}) and which has the same behavior as
$C_2(J_1,J_2;T)$ for $T\sim T_{2c}$ as shown in (\ref{lowersc}).
We have also defined in Part I \cite{HAYMEF} the net
{\it enhancement} of the specific heat as
\ba
&&{\cal E}(J_1,J_2;m_1,m_2; T)\nonumber\\
&&\quad=C(J_1,J_2; T)-C(J_1,0; T)-C{\bf (}0,J_2; {\checkT}(r){\bf)}.
\label{enhancement}\ea
Near $T_{1c}$, we find that $C(0,J_2; {\checkT})$ is similar
to $C(0,J_2; T)$ in that it is relatively small and nonsingular
and, in fact, does not contribute to the scaling function.
We may use (\ref{C1}), (\ref{finites}) and (\ref{finite1}) to
rewrite the enhancement as
\be
{\cal E}(J_1,J_2;m_1,m_2; T)=\frac {C_1(T)-C^\infty(T)}{1+s}+
\delta C,\label{enh1}\ee
where we define the difference $\delta C$ as
\be
\delta C=C(0,J_2; T)-C{\bf (}0,J_2; {\checkT}(r){\bf)}\simeq 0.
\ee
Indeed for $\e^{-2m_2/\xi_2(T)}\ll 1$ we find that $C_1(T)$ has
the same scaling behavior as $C^\infty(T)$. From (\ref{uppersc}),
we thus find that the enhancement is of the order of a
{\it correction to scaling}. As (\ref{sigma1}) gives the magnitude
of the corrections to scaling, we find that (\ref{enh1}) becomes
\be
{\cal E}(J_1,J_2;m_1,m_2; T)\approx
\frac {B_0(r)\ln m_1+B(r,\tau_1)}{m_1+m_2},
\label{upperenh}\ee 
where $B_0(r)$ and $B(r,\tau_1)$ are functions of order unity
whose forms can be gauged from Figs.\ 9 to 11 of Part I. On the
other hand we find from (\ref{lowersc}) the corresponding result
\be
{\cal E}(J_1,J_2;m_1,m_2; T)\approx
\frac {{\hat B}_0(r)\ln m_2+{\hat B}(r,\tau_2)}{m_1+m_2},
\label{lowerenh}\ee 
for $T$ near $T_{2c}$, when $\e^{-2m_1/\xi_1(T)}\ll 1$,
with ${\hat B}_0(r)$ and ${\hat B}(r,\tau_2)$ appropriate functions
of order unity. As the relative strength $r$ increases, $T_{2c}$
and $T_c$ approach $T_{1c}$, because $T_{2c}=rT_{1c}$ and
$T_{2c}<T_c<T_{1c}$. This also mean that the regimes in which
(\ref{uppersc}) or (\ref{lowersc}) are valid shrink.
The explicit form of these corrections to scaling and the
functions $B_0(r)$, $B(r,\tau_1)$, ${\hat B}_0(r)$ and
${\hat B}(r,\tau_2)$ are not easy to obtain exactly and the
computations have not been attempted. 

%%%%%%%%%%%%%%%%%%%%%%%%%%%%%%%%%%%%%%%%%%%%%%%%%%%%%%%%%%
\section{Summary}
For the alternating layered Ising model, we show there exists a
well defined critical temperature, at which, the specific heat
diverges according to (\ref{log}). However, for fixed relative
strength $r=J_2/J_1\ne 1$, and $s=m_2/m_1\ne 0$, we find the
amplitude $A(r,s)$ decreases exponentially fast in $m_2$. For
large enough $m_1$ and $m_2$ the specific heat also has two distinct
maxima satisfying the relations $T_c<T_{max1}<T_{1c}$ and
$T_{2c}<T_{max2}<T_{c}$. These general results agree with the
experiments on superfluid helium by Gasparini and coworkers.
\cite{GKMD,PKMG,PKMGn,MEF,PG,PKMGpr}

Near $T_{1c}$, we find the net specific heat, $C_1(T)$ defined
in (\ref{C1}), obeys finite-size scaling as established in
(\ref{uppersc}) when $\e^{-2m_2/\xi_2(T)}$ is negligible. On the
other hand, near $T_{2c}$, the lower maximum, we find the
corresponding $C_2(T)$, whose free energy is defined in (\ref{f2}),
obeys the finite-size scaling given by (\ref{lowersc}) when
$\e^{-2m_1/\xi_1(T)}$ is small; remarkably, the sign of the
appropriate scaled temperature deviation, $T-T_{2c}$ is then
{\it reversed} from that for an infinite strip of finite width.
However, this corresponds qualitatively to the observed enhancement
in the experiments induced by the proximity effects of ordered
regions below the true bulk critical point.

It should be remarked, however, that modelling the experimental
systems would be improved by using three spatial dimensions and,
furthermore, Ising spins would better be replaced by XY spins.

%%%%%%%%%%%%%%%%%%%%%%%%%%%%%%%%%%%%%%%%%%%%%%%%%%%%%%%%%%
\begin{acknowledgments}
The author would like to thank Professor M.~E.\ Fisher for
suggesting the problem. His valuable comments and criticism on
the manuscript are deeply appreciated. Thoughtful and helpful
comments on the manuscript by Professor F.~M.\ Gasparini are
also gratefully acknowledged.
This work was supported in part by the National Science Foundation
under grant No.\ PHY-07-58139.
\end{acknowledgments}

%%%%%%%%%%%%%%%%%%%%%%%%%%%%%%%%%%%%%%%%%%%%%%%%%%%%%%%%%%
%\section*{References}%


\begin{thebibliography}{000}
\bibitem{note1}{ This error is due to the term in the free energy
given by (1.22) in Ref.~\onlinecite{MEFJp} where the expression
$t\tan(1/t)\to|t|\pi/2$ had been used for the limit $t\to0$. If
however, one evaluates the second derivative of this term first
and then taken the limit $t\to0$, one finds that its contribution
is of the relative order $1/m$.}
\bibitem{note2}{Because of $K_2=rK_1$, the second derivative with
respect to $K_1$ in (\ref{spheat}) picks up the factor $r^2$.
However, this factor is canceled out by evaluating $K^2_1$ in
(\ref{spheat}) at $T=T_{2c}$. Thus the scaling function is
universal, independent of the value of $J$.}
\end{thebibliography}
\end{document}